\documentclass[a4paper,12pt]{article}

\usepackage{monash}
\usepackage{relaxion}

\bibliography{relaxion}

\begin{document}

\title{Naturalness of the relaxion mechanism}

\author{Andrew Fowlie}
\author{Csaba Balazs}
\author{Graham White}
\affiliation{ARC Centre of Excellence for Particle Physics, School of Physics and Astronomy, Monash University, Melbourne, Victoria 3800 Australia}
\author{Luca Marzola}
\author{Martti Raidal}
\affiliation{National Institute of Chemical Physics and Biophysics, R\"avala 10, Tallinn 10143, Estonia}
\affiliation{Institute of Physics, University of Tartu, Ravila 14c, Tartu 50411, Estonia}

\date{\today} 

\maketitle

\begin{abstract}
The relaxion mechanism is a novel solution to the hierarchy problem.
In this first statistical analysis of the relaxion mechanism, we quantify the 
relative plausibility of a QCD and a non-QCD relaxion model
versus the Standard Model with Bayesian statistics, 
which includes an automatic penalty for fine-tuning.
We find that in light of the hierarchy between the weak and Planck scales, relaxion models are 
favoured by colossal Bayes-factors.
Constraints upon \eg the vacuum energy
during relaxation, however, shrink the Bayes-factors such that relaxion models are only slightly favoured. 
Including the bounds on $\thetaQCD$ shatters the plausibility of the QCD relaxion 
model as it typically yields $\thetaQCD \gg 0$. 
Finally, we augment our models with scalar-field inflation and consider 
measurements of inflationary observables from BICEP/Planck. 
We find that, all told, the Standard Model is favoured by huge Bayes-factors as
the relaxion models require fine-tuning such that 
the Hubble parameter is less than the height of the periodic barriers.
Thus, whilst we confirm that relaxion models could solve the hierarchy problem, 
we find that their unconventional cosmology demolishes their plausibility.
\end{abstract}

\section{Introduction}

Graham \etal\cite{Graham:2015cka} recently proposed a relaxation mechanism\cite{Abbott:1984qf,Dvali:2003br,Dvali:2004tma} that solves
the hierarchy problem\cite{Weinberg:1975gm,Weinberg:1979bn,Susskind:1978ms,Gildener:1976ai}
by utilising the dynamics of an
axion-like field, dubbed the relaxion. In the Standard Model (SM), the hierarchy problem originates from quadratic corrections
to the weak scale. Whereas supersymmetry
cancels them with new quadratic corrections involving
supersymmetric particles\cite{Witten:1981nf}, the relaxion mechanism cancels them with the vacuum expectation value (VEV) of a relaxion field.

The ingenuity of the relaxion mechanism is that the dynamics of the relaxion field ensure a precise cancellation 
without patent fine-tuning of parameters or initial conditions. Within the relaxion paradigm, 
interactions between a complex Higgs doublet, $h$, and an axion-like field, $\phi$, govern the 
weak scale via the scalar potential\cite{Graham:2015cka}
\begin{equation}\label{Eq:Relaxion}
V = \left(\mu^2 - \kappa \vev{a} \phi \right) h^2 - m_b^3 \vev{h} \cos\left( \frac{\phi}{f}\right) - m^2 \vev{a} \phi + \lambda h^4,
\end{equation}
where, because of quadratic corrections, we expect that the masses should be close to the cut-off $\Lambda$, \ie $\mu^2\sim m^2 \sim \Lambda^2$, 
$m_b$ and $f$ are coupling constants of dimension mass, \vev{a} is the VEV of a spurion field that breaks a shift symmetry $\phi \to \phi + 2\pi f$, 
$\kappa$ is a dimensionless coupling, and \vev{h} is the VEV of the Higgs field, which is a function of the 
relaxion field $\phi$.

Let us label the co-efficient of $h^2$ in the relaxion potential 
\begin{equation}\label{Eq:H_lit}
m_h^2(\phi) \equiv \mu^2 - \kappa \vev{a} \phi,
\end{equation}
for convenience, such that the Higgs VEV may be written
\begin{equation}
\vev{h} = 
\begin{cases} 
      \sqrt{\frac{-m_h^2(\phi)}{2\lambda}} & m_h^2(\phi) < 0,\\
      0 & \text{otherwise.}
\end{cases}
\end{equation}
If the Higgs VEV is non-zero, the 
cosine term provides a periodic barrier for the 
relaxion field with barriers separated by $2\pi f$. In the unbroken
phase in which $\vev{h}=0$, the barrier is down and the relaxion field slowly rolls down a linear potential.
Once $m_h^2(\phi) < 0$, however, the potential is such that the Higgs field acquires a VEV, $\vev{h}\neq0$, 
breaking electroweak symmetry (EWSB) and raising the periodic barrier. The now-raised periodic barrier traps the
relaxion field in a minimum. If the relaxion field cannot roll past a local minimum, it results in a weak scale of about  
\begin{equation}
\vev{h} \gtrsim f \frac{m^2 \vev{a}}{m_b^3}.
\end{equation}
Thus this mechanism could result in $\vev{h} \ll \MP$.

We require, \latin{inter alia}, that the relaxion field
dissipates energy as it rolls or else it would have sufficient kinetic energy
to surmount the periodic barriers. In the relaxion paradigm, this is ensured by Hubble friction --- 
a term analogous to a friction term in the Euler-Lagrange equation for the 
relaxion field originating from the expansion of the Universe \see{Mukhanov:2005sc}:
\begin{equation}
\ddot \phi + 3 H \dot \phi + \frac{\partial V}{\partial \phi} = 0,
\end{equation}
where $H$ is the Hubble parameter. If Hubble friction is substantial, the relaxion
field could be in a slow-roll regime in which the acceleration
$\ddot \phi$ can be neglected. 

Ostensibly, the relaxion mechanism ameliorates fine-tuning associated with
the weak scale, but Raidal \etal\cite{DiChiara:2015euo}  
stress that it could require a fine-tuned inflationary sector if the relaxion
is the QCD axion.
Unfortunately, there is no consensus in high-energy
physics on the appropriate measure of fine-tuning or about the logical foundations of fine-tuning arguments, despite their prominence. 
In earlier work to judge fine-tuning in relaxion models, Jaeckel \etal\cite{Jaeckel:2015txa} developed a
new formalism based on their intuition, whereas Raidal \etal\cite{DiChiara:2015euo} utilised
common Barbieri-Giudice style measures\cite{Barbieri:1987fn,Ellis:1986yg}. In \refsec{Sec:Bayes},
we critique Jaeckel's approach and instead advocate a Bayesian methodology, 
discussed numerous times over the last decade in the context of fine-tuning in supersymmetric models\cite{Allanach:2007qk,Cabrera:2008tj,Fowlie:2014xha,Fichet:2012sn,Cabrera:2010dh,Fowlie:2015uga,Kim:2013uxa,Fowlie:2014faa}.
In this methodology, in light of experimental data about the weak scale and inflation, we update our belief in a model with a Bayesian evidence. We further analyse
the relaxion potential in \refsec{Sec:Analysis}. We describe our models --- minimal relaxion models and the SM augmented by
scalar-field inflation --- in \refsec{Sec:Models} and calculate their Bayesian evidences in \refsec{Sec:Results}. 
This is the first statistical analysis of a relaxion model. We close in \refsec{Sec:Conclusions} with a brief discussion of our findings.

\section{Bayesian fine-tuning}\label{Sec:Bayes}
Bayesian statistics provides a logical framework for updating 
beliefs in scientific theories in light of data \see{Earman,Jaynes:2003,Gregory}. 
This methodology is becoming increasingly common in high-energy physics
\see{LopezFogliani:2009np,Kowalska:2012gs,Williams:2015bfa,Diamanti:2015kma,Catalan:2015cna,
Casas:2014eca,Roszkowski:2014wqa,Strege:2014ija,Fowlie:2014awa,AbdusSalam:2013qba,Cabrera:2013jya,
Arina:2013zca,deAustri:2013saa,Fowlie:2013oua,
Cabrera:2012vu,Strege:2012bt,Balazs:2012bx,Fowlie:2012im,
Balazs:2013qva,Roszkowski:2012uf,Strege:2011pk,Fowlie:2011mb,
Bertone:2011pq,Cabrera:2011ds,Allanach:2011ya,Bertone:2011nj,Cumberbatch:2011jp,
Fowlie:2011vf,Allanach:2011ut,Feroz:2011bj,Ripken:2010ja,Cabrera:2010xx,Akrami:2010cz,
Cabrera:2009dm,Akrami:2009hp,Roszkowski:2009ye,Roszkowski:2009sm,Trotta:2008bp,
Feroz:2008wr,Allanach:2008iq,Allanach:2008tu,Allanach:2008zn,Roszkowski:2007fd,Roszkowski:2006mi}
and cosmology \see{Trotta:2008qt,Martin:2013nzq,Martin:2014lra}, and arguably
captures the essence of the hierarchy problem\cite{Allanach:2007qk,Cabrera:2008tj,Fowlie:2014xha,Fichet:2012sn,Cabrera:2010dh,Fowlie:2015uga,Kim:2013uxa,Fowlie:2014faa} and 
the principle of Occam's razor \see{1991BAAS...23.1259J}. We briefly recapitulate the essential details.

The Bayesian framework enables one to assign numerical measures to degrees of belief. 
To assess two models, $\model_a$ and $\model_b$, one begins by quantifying one's relative degree of belief in the models,
prior to considering any experimental data. This is known as the prior odds,
\begin{equation}
\text{Prior odds} \equiv \frac{\P{\model_a}}{\P{\model_b}},
\end{equation}
where $\P{\model}$ is one's prior belief in a model \model.
From the prior odds, we can calculate the posterior odds --- one's relative degree of belief in the models updated with experimental data,
\begin{equation}
\text{Posterior odds} \equiv \frac{\Pg{\model_a}{\data}}{\Pg{\model_b}{\data}},
\end{equation}
where \data represents experimental data \eg in this work data from BICEP/Planck.
The prior odds and the posterior odds are related by a so-called Bayes-factor:
\begin{equation}
\text{Posterior odds} = \text{Bayes-factor} \times \text{Prior odds}.    
\end{equation}
By applying Bayes' theorem, it can be readily shown that the Bayes-factor is a ratio of probability densities,
\begin{equation}
\text{Bayes-factor} \equiv \frac{\pg{\data}{\model_a}}{\pg{\data}{\model_b}},
\end{equation}
where the probability densities in question are known as Bayesian evidences or just evidences. 
The evidence for a model \model can be calculated by Bayes' theorem and marginalisation,
\begin{equation}\label{Eq:Evidence}
\ev \equiv \pg{\data}{\model} = \int \pg{\data}{\model, \params} \cdot \pg{\params}{\model} \prod \dif \params
\end{equation}
where $\params$ are the model's parameters, $\pg{\data}{\model, \params}$ is a so-called likelihood function ---
the probability density of our observed data given parameters $\params$ --- and  \pg{\params}{\model} is our prior density
for the model's parameters \params.

The likelihood function is uncontroversial as its form is dictated by the 
nature of an experiment and it is a critical ingredient
in Bayesian and frequentist statistics. The role and form of the prior density, however, 
remain contentious issues.  In as much as it is possible, we pick objective
priors that reflect our knowledge or ignorance about a parameter and respect
rational constraints from \eg symmetries.

We calculate Bayes-factors for the SM augmented with scalar-field inflation (\SMsigma) versus relaxion models. The final step ---
that of updating one's prior odds with a Bayes-factor to find one's posterior odds --- 
is left to the reader. That is not to say that a Bayes-factor is independent of 
any prior choices --- it is in fact a functional of the
priors for the parameters of the  models in question.

Before closing, we briefly discuss attempts to quantify fine-tuning in a relaxion
model by Jaeckel \etal\cite{Jaeckel:2015txa} and by Raidal \etal\cite{DiChiara:2015euo}. Raidal \etal
employed Barbieri-Giudice style measures of fine-tuning\cite{Barbieri:1987fn,Ellis:1986yg}. 
Whilst intuitive, such measures lack a logical foundation, though emerge in intermediate
steps in a calculation of the Bayesian evidence\cite{Allanach:2007qk,Cabrera:2008tj,Fowlie:2014xha,Fichet:2012sn,Cabrera:2010dh,Fowlie:2015uga,Kim:2013uxa,Fowlie:2014faa}. 
Jaeckel \etal developed a novel measure of electroweak fine-tuning, $F$, based on
the fraction of a model's parameter space, \params, that predicts a weak scale less than that
observed:
\begin{equation}\label{Eq:Jaeckel}
\frac1F \equiv \frac{V_{v(\params) \le v}}{V} = \frac{\int \theta(v - v(\params)) \prod \dif \params}{\int \prod \dif \params}.
\end{equation}
This measure contrasts with Barbieri-Giudice measures in that it considers a model's 
entire parameter space rather than a single point in it.
Jaeckel's measure, however, depends on one's choice of 
parameterisation or measure for the parameter space.

Curiously, Jaeckel's measure in \refeq{Eq:Jaeckel} is reminiscent of the Bayesian evidence if one
considers measurements of the weak scale, especially if one writes (unnecessary) normalisation factors for the priors,
\begin{equation}
\ev = \frac{\int \pg{v}{\model, \params} \cdot \pg{\params}{\model} \prod \dif \params}{\int \pg{\params}{\model} \prod \dif \params}
\quad\text{vs.}\quad
\frac1F = \frac{\int \theta(v - v(\params)) \prod \dif \params}{\int \prod \dif \params}.
\end{equation}
The differences are that Jaeckel \etal pick a step-function for the likelihood for the weak scale, $v$, 
and omit a measure for the volume of parameter space, \ie
a prior. In other words, by following their noses and attempting to formulate
fine-tuning in a logical manner, Jaeckel \etal create
an ersatz Bayesian evidence, though fail to recognise the dependence of their
fine-tuning measure upon the measure assigned to the parameter space.

\section{Analysis of relaxion potential}\label{Sec:Analysis}
\begin{figure}[ht]
\centering
\includegraphics[width=0.99\textwidth]{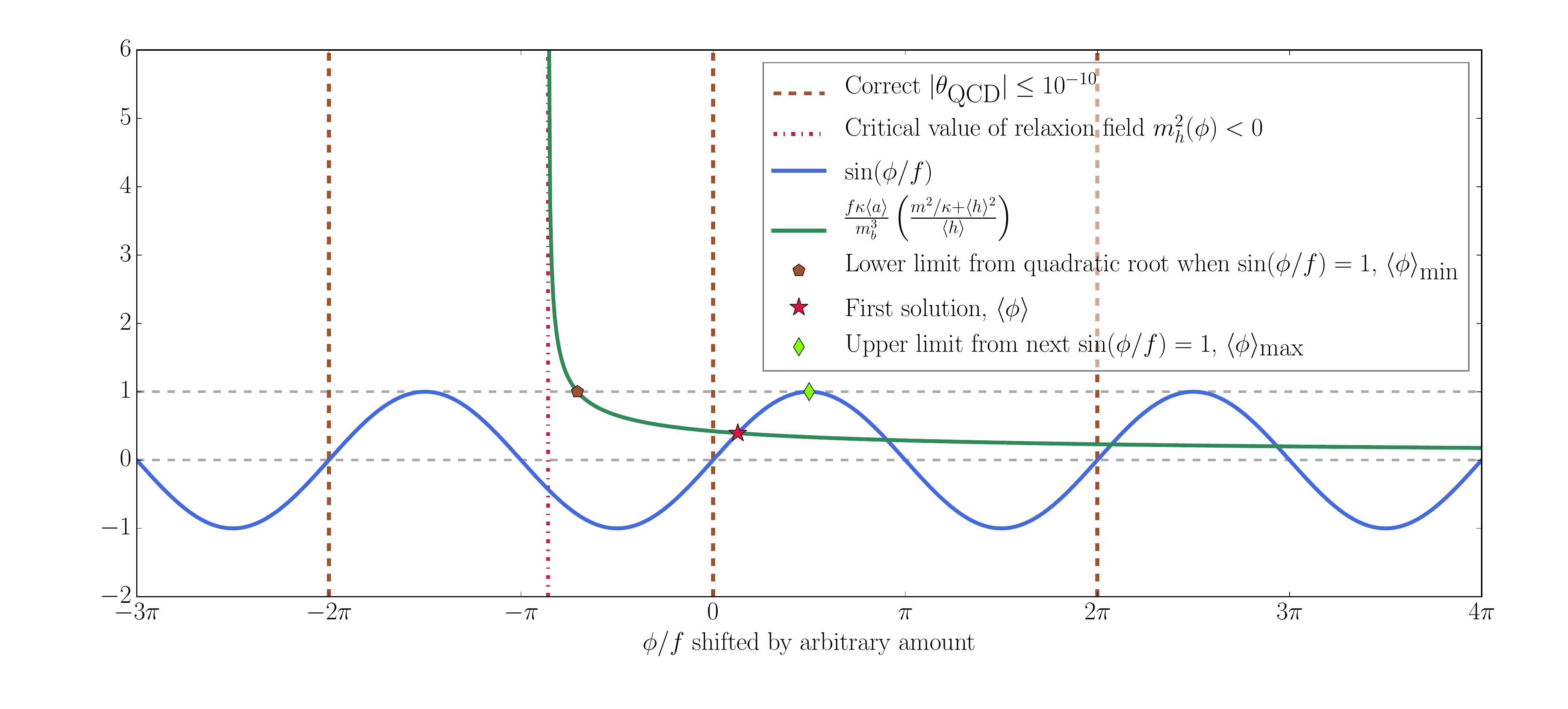}
\caption{Graphing the left-hand side (blue line) and right-hand side (green line) of the transcendental equation in \refeq{Eq:Transcendental}.
The solution (red star) lies in the interval in \refeq{Eq:int}, marked by a brown pentagon and a green diamond. 
In unusual cases, the second point at which the right-hand side equals plus one (not shown) may be a sharper bound. 
The $\phi/f$-axis is shifted such that correct \thetaQCD occurs at small multiples of $2\pi$ (vertical brown dashed lines) close to the solution. 
EWSB is broken once the critical value of the relaxion field is surpassed (vertical red dot-dashed line).}
\label{Fig:Graphing}
\end{figure}

Let us further analyse the relaxion potential in \refeq{Eq:Relaxion},
\begin{equation}
V = \left(\mu^2 - \kappa \vev{a} \phi \right) h^2 - m_b^3 \vev{h} \cos\left( \frac{\phi}{f}\right) - m^2 \vev{a} \phi + \lambda h^4.
\nonumber
\end{equation}
As in \refcite{Espinosa:2015eda}, for simplicity we consider only linear terms in the relaxion field $\phi$. 
The equations $\partial V / \partial \phi = 0$ and $\partial V / \partial h = 0$ result in a transcendental equation,
\begin{equation}\label{Eq:Transcendental}
\sin(\phi/f) = \frac{f \kappa \vev{a}}{m_b^3} \left(\frac{m^2/\kappa + \vev{h}^2}{\vev{h}} \right).
\end{equation}
By graphing as in \reffig{Fig:Graphing}, one finds that if there is a solution, it lies in the interval $\vev{h}_\text{min} \leq \vev{h} \leq \vev{h}_\text{max}$ where
\begin{equation}\label{Eq:int}
\vev{h}_\text{min} = \frac{m_b^3 - \sqrt{m_b^6 - 4 \kappa m^2 \vev{a}^2 f^2}}{2 \kappa \vev{a} f} 
\quad\text{and}\quad 
\vev{h}_\text{max} = \sqrt{\frac{-\mu^2 + \kappa \vev{a} \vev{\phi}_\text{max}}{2\lambda}},
\end{equation}
and
\begin{equation}
\vev{\phi}_\text{min} = \frac{2\lambda \vev{h}_\text{min}^2 + \mu^2}{\kappa \vev{a}}
\quad\text{and}\quad
\vev{\phi}_\text{max} = (2n + 1/2)\pi f
\end{equation}
where $n$ is the smallest integer such that $\vev{\phi}_\text{max} > \vev{\phi}_\text{min}$. 
If the square-root is imaginary, there are no solutions, otherwise, there are zero to four solutions inside the interval, which must be identified numerically. 
The interval results from recognising that a solution must lie between the point at which the right-hand side of \refeq{Eq:Transcendental} equals plus one, matching the maximum of the left-hand side, 
and the subsequent point at which the latter is again maximal.
If required, one can improve this interval with piece-wise expressions by graphing. In some cases, the positive quadratic root,
similar to that for $\vev{h}_\text{min}$, is a sharper bound for $\vev{h}_\text{max}$. If the barrier height is substantial, 
the lower bound reduces to the approximation for $\vev{h}$ in \refeq{Eq:H_lit}, that is, 
\begin{equation}\label{Eq:Approx}
\vev{h}_\text{min} \approx f \frac{m^2 \vev{a}}{m_b^3} \quad\text{if}\quad \frac{4 \kappa m^2 \vev{a}^2 f^2}{m_b^6} \ll 1.
\end{equation}
This implies that $\kappa \vev{a} \ll m_b^3 / (4 f \vev{h})$. 
A necessary (though not sufficient) condition for solutions to the transcendental equation is that the root in \refeq{Eq:int} must be real,
\begin{equation}
\frac{4 \kappa m^2 \vev{a}^2 f^2}{m_b^6} \le 1.
\end{equation}

In other words,
the relaxion mechanism ensures that the weak scale is independent of quadratic 
corrections to the Higgs mass from a cut-off or unknown high-scale physics, solving the hierarchy problem. In fact,
the Higgs VEV is bounded by an expression that is independent of the Higgs mass,
\begin{equation}\label{Eq:HLimits}
\vev{h}_\text{min} \le \vev{h} \le \vev{h}_\text{max} \leq \sqrt{\vev{h}_\text{min}^2 + \kappa \vev{a} f \pi/\lambda}.
\end{equation}
The Higgs mass $\mu^2$ and any corrections affect the position of \vev{h} inside this interval, but not the interval
itself. The width of this interval is typically small such that numerically solving for the VEV of the Higgs field inside the interval may be unnecessary. 

Unfortunately, if the
relaxion is the QCD axion, we expect  that barrier height
$m_b^3$ is connected to QCD, such that $m_b \sim \Lambda_\text{QCD}$, resulting in
\begin{equation}
\vev{a} \lesssim \vev{h}  \times 10^{-18} \simeq 10^{-16} \gev,
\end{equation}
where we impose an experimental limit on the QCD decay constant, $f \gtrsim 10^9 \gev$, and pick $m \simeq 1 \tev$ in 
\refeq{Eq:Approx}. Thus
achieving a
small weak scale requires a tiny spurion VEV, $\vev{a} \lll \MP$. Such a small coupling may be
natural as it breaks a shift-symmetry \see{'tHooft:1979bh}; however, there may
be issues due to the gauge symmetry at the basis of the construction\cite{Gupta:2015uea,Ibanez:2015fcv}.

\subsection{\thetaQCD in relaxion models}\label{Sec:ThetaQCD}

Let us investigate whether a relaxion model might resolve the strong-CP problem\cite{Peccei:1977hh}
by explaining $\thetaQCD \lesssim 10^{-10}$\cite{Agashe:2014kda}. \latin{Prima facie}, the expression for
\thetaQCD is simple \see{Donoghue:1995va},
\begin{equation}\label{Eq:SimpleThetaQCD}
\thetaQCD = \left| \vev{\phi} / f 
\quad\text{on $-\pi$ to $\pi$} \right|.
\end{equation}
Numerically, however, this cannot be used for 
calculating \thetaQCD\ --- as $\vev{\phi}/f$ is substantial, there is a breakdown in numerical precision in expressions
such as $\vev{\phi}/f \mod 2\pi$. Instead, we find the principal solution for \thetaQCD,
\begin{equation}
\thetaQCD = \arcsin \left|\frac{f \kappa \vev{a}}{m_b^3} \left( \frac{m^2/\kappa + \vev{h}^2}{\vev{h}} \right)\right|,
\end{equation}
by utilising \refeq{Eq:Transcendental}.

The minimum \thetaQCD obtainable occurs at the minimum of the right-hand side of \refeq{Eq:Transcendental}, such that, if there is 
a solution at that point, $\sin(\phi/f)$ is as close to zero as possible. Thus we find that
\begin{equation}\label{Eq:ThetaQCD}
\min \thetaQCD = 
\begin{cases} 
      \arcsin \left|\frac{2 \vev{a} f m \sqrt{\kappa}}{m_b^3} \right| & \text{if $\vev{h}_\text{minima} \le \vev{h}_\text{max}$},\\
      \arcsin \left|1 - \frac{\kappa \vev{a} f\pi}{2\lambda \vev{h}_\text{min}^2} + \cdots \right| = \pi/2 - \sqrt{2\epsilon} + \cdots
      & \text{otherwise,}
\end{cases}
\end{equation}
where $\vev{h}_\text{minima} = m / \sqrt{\kappa}$ which minimises the right-hand side of
\refeq{Eq:Transcendental} and the second line is never less than the first line. 
The terms represented by the ellipses are higher powers of $\eta$ and $\epsilon$, 
\begin{equation}\label{Eq:ThetaQCDApprox}
\vev{h}_\text{max} \approx \vev{h}_\text{min}(1+\epsilon)\quad\text{where}\quad\epsilon \equiv \frac{\kappa \vev{a} f\pi}{2\lambda \vev{h}_\text{min}^2} \ll 1 \quad \text{and} \quad \eta \equiv \frac{\kappa \vev{h}_\text{min}^2}{m^2}\ll 1.
\end{equation}
We expand to first order in $\epsilon$ and neglect all powers of $\eta$. 
\refeq{Eq:ThetaQCD} originates from considering that the minimum possible \thetaQCD would 
occur at the minimum of the right-hand side of \refeq{Eq:Transcendental} if that minima occurred at $\phi /f \approx 2n\pi$ and not $\phi /f \approx (2n+1)\pi$, 
such that $\thetaQCD \approx 0$ and not $\thetaQCD \approx \pi$. 
If that minima occurs, however, outside the interval for the possible solutions for $\vev{h}$, 
it is impossible. In that case, 
the right-hand side of \refeq{Eq:Transcendental} is monotonic inside the interval for the possible solutions for $\vev{h}$ and we consider the 
right-hand side evaluated at $\vev{h}_\text{max}$ from \refeq{Eq:HLimits}. 
As the general expression is rather complicated, we apply the approximations in \refeq{Eq:ThetaQCDApprox}, 
which are reasonable for phenomenologically viable points. 
In fact, phenomenologically viable points are always in the second regime 
in which $\vev{h}_\text{minima} = m / \sqrt{\kappa} \gg \vev{h}_\text{max}$, 
such that $\min \thetaQCD \approx \arcsin 1 = \pi/2$. This is confirmed in our numerical analysis.

\subsection{Finite-temperature effects}

In this paper and in the literature so far,
the relaxion mechanism was analysed at zero temperature.\footnote{We note, however, that \refcite{Hardy:2015laa}
considers finite-temperature effects in an alternative relaxion mechanism.} 
Finite-temperature effects could, however, non-trivially 
affect the relaxion potential in \refeq{Eq:Relaxion}:
\begin{itemize}

\item Non-perturbative effects responsible for the induced 
effective potential of the relaxion are temperature 
dependent\cite{Gross:1980br}. 
However, since this affects only the heights of the barriers 
and not their spacing and since Hubble friction during 
inflation is typically substantial, 
it is unclear whether finite-temperature effects would 
impact the viability of the relaxion mechanism.  

\item Finite-temperature corrections to the effective 
potential would alter the shape of the potential 
(increasing the gradients of the slopes), possibly
delaying the onset of EWSB\cite{Quiros:1999jp}. 
If EWSB is delayed until a late time (corresponding to a lower temperature after reheating), 
it could constrain when inflation must start through the requirement that it lasts at least $50$ $e$-folds after EWSB. 
Furthermore, the flat regions of the zero-temperature 
inflaton potential are strongly modified by finite-temperature 
effects. 

\item We find that the reheating temperature in our relaxion models is typically of order $10^{10}\gev$.
At such a high temperature, electroweak
symmetry could be easily restored, with the effect of further hindering the viability 
of the model. 
\end{itemize}
Clearly all the mentioned finite-temperature effects have the potential to impose further constraints on
the relaxion model parameter space, to an extent that will be quantified in following projects.

\subsection{Baryon asymmetry}

We observe a significant baryon asymmetry in our Universe. 
Sakharov\cite{Sakharov:1967dj} demonstrated that 
generating this asymmetry --- baryogenesis --- would require 
a departure from thermal equilibrium, $\mathcal{C}$ and 
$\mathcal{CP}$ violation, and baryon number violation. 
In the relaxion paradigm, however, the final $50$ or so
 $e$-folds of inflation occur during or immediately after 
 EWSB and inevitably wash-out any potential net baryon 
 number generated in this process \see{Riotto:1998bt}.
Novel mechanisms that invoke multi-step phase transitions are
 also ruled out since the fields must be in the final SM 
 vacuum at the end of inflation.\footnote{One could in principle have a multi-step phase
  transition that departed from the SM vacuum for 
  baryogenesis and later returned to it but this 
  somewhat undermines the motivation for relaxion models.}
Scenarios in which the inflaton itself could generate 
the required baryon asymmetry 
\see{Rangarajan:2001yu,Alexander:2004us} also appear to 
be incompatible with the relaxion mechanism because of the 
further constraints implied by the already heavily
constrained dynamics of the inflaton.
Finally weak sphalerons are also exponentially
VEV suppressed\cite{Patel:2011th} after the
electroweak phase transition which means that any subsequent baryogenesis scenario would have to 
rely on a different source of baryon and lepton number violation.
As we shall see, these difficulties would strengthen our 
conclusions about the viability of inflation in the present 
framework. 

\section{Description of models}\label{Sec:Models}

We apply Bayesian model comparison to three models: the SM augmented with single-field
scalar-field inflation (\SMsigma), a QCD relaxion model and a general relaxion model. 
For other relaxion models, see \eg \refcite{Matsedonskyi:2015xta,Kaplan:2015fuy,Choi:2015fiu,Espinosa:2015eda,Antipin:2015jia,Batell:2015fma,Fonseca:2016eoo}.
Ultimately, we wish to find whether the 
relaxion mechanism ensures that a relaxion model is favoured by the Bayesian evidence
versus the SM\@. In each model, all scalar-fields receive quadratic corrections to their masses
from a cut-off, $\Lambda$, which lies close to the Planck scale.

\subsection{The Standard Model with scalar-field inflation}

The SM Higgs sector is described by two bare Lagrangian parameters --- $\mu^2$ and $\lambda$ --- in the SM Higgs potential,
\begin{equation}
V_h = \mu^2 |h|^2 + \lambda |h|^4,
\end{equation}
and a cut-off at which the bare parameters are specified, $\Lambda$. We augment
the SM with mixed inflation, a canonical 
model of scalar-field inflation \see{Martin:2013tda}. Mixed inflation is described by an
inflaton mass, $m_\sigma^2$, and quartic coupling, $\lambda_\sigma$, in a potential
\begin{equation}
V_\sigma = \frac12 m_\sigma^2 \sigma^2 + \lambda_\sigma \sigma^4
\end{equation} 
and the number of $e$-folds, \Nefold. We denote this model by \SMsigma.

Note that in the \SMsigma model, the evidence approximately factorises into a factor for the weak scale and a
factor for the inflationary observables, $r$, $n_s$ and $A_s$,
\begin{align}
\begin{split}
\ev ={}& \pg{\mz, r, n_s, A_s}{\SMsigma}\\
    \approx{}& \pg{\mz}{\SMsigma} \cdot \pg{r, n_s, A_s}{\SMsigma}\\
    ={}& \pg{\mz}{\SM} \cdot \pg{r, n_s, A_s}{\sigma}  
\end{split}
\end{align}
as the measurements are independent and model parameters that affect inflationary observables do not affect
the weak scale and \latin{vice-versa}, with the exception of the cut-off, $\Lambda$, 
which results in quadratic corrections to the inflaton mass and the Higgs mass.

\subsubsection{Calculation of observables}
For comparison with measurements from Planck in our statistical analysis, 
we calculate inflationary observables via the so-called slow-roll parameters\cite{Liddle:1992wi}:
\begin{equation}
\epsilon(\sigma) \equiv \frac{\MP^2}{16\pi} \left(\frac{V^\prime(\sigma)}{V(\sigma)}\right)^2
\quad\text{and}\quad
\eta(\sigma) \equiv \frac{\MP^2}{8\pi} \frac{V^{\prime\prime}(\sigma)}{V(\sigma)}
\end{equation}
where a prime indicates a derivative with respect to the inflaton field $\sigma$.
Inflation finishes once the inflaton field reaches a value $\sigma_f$ such that $\epsilon(\sigma_f) = 1$. 
The number of $e$-folds desired before inflation ends (and in the case of the relaxion, after EWSB),
\begin{equation}
\Nefold \simeq \frac{-8\pi}{\MP^2} \int_{\sigma_i}^{\sigma_f} \frac{V(\sigma)}{V^\prime(\sigma)} \dif \sigma,
\end{equation}
determines the inflaton field at the beginning of inflation, $\sigma_i$. 
The number of $e$-folds desired should be $\Nefold\gtrsim50$. 
The spectral index, $n_s$, and the ratio of scalar to tensor perturbations, $r$, may be 
written to first order in the slow-roll parameters as \see{Liddle:2000cg}
\begin{equation}
n_s = 1 - 6\epsilon(\sigma_i) + 2\eta(\sigma_i)
\quad\text{and}\quad
r = 4\pi\epsilon(\sigma_i).
\end{equation}
The normalisation of the potential governs the
amplitude of scalar perturbations and the Hubble parameter,
\begin{align}
A_s &= \frac{1}{\MP^6} \frac{128 \pi}{3} \frac{V(\sigma_i)^3}{|V^\prime(\sigma_i)|^2},\\
H & = \sqrt{\frac{V(\sigma_i)}{3 \MP^2}},
\end{align}
but cannot affect $r$ or $n_s$. The normalisation of the scalar perturbations 
is arbitrary and varies in the literature. For comparison with Planck data, we pick that of the Planck experiment \see{Planck:2013jfk}. 
We include a quadratic correction to the inflaton
mass --- to include a dominant quantum contribution to fine-tuning --- but otherwise
our formulas are tree-level. We solve for the inflaton field at the beginning and end
of inflation, $\sigma_i$ and $\sigma_f$, with numerical methods.

The mass of the $Z$ boson --- which represents the weak scale --- is calculated in 
the usual manner,
\begin{equation}
\mz^2 = \frac{-g^2}{2\lambda} \left(\mu^2 + \beta\Lambda^2\right),
\end{equation}
where $\beta$ is a loop factor. The QCD phase is an input parameter.

\subsection{Relaxion models}

We consider two relaxion models described by the potential in \refeq{Eq:Relaxion}. 
In the first model, we do not identify the relaxion with the Peccei-Quinn axion that solves the 
strong CP problem\cite{Peccei:1977hh}, whereas in the second model, the relaxion is indeed the Peccei-Quinn axion.  

For a necessary epoch of low-scale inflation after the relaxion mechanism, we extend the relaxion potential in \refeq{Eq:Relaxion} 
by the most general renormalisable single-field inflaton potential \see{DiChiara:2015euo} with an inflaton field $\sigma$,
\begin{equation}
V = m_3^3 \sigma + \frac12 m_2^2 \sigma^2 + \frac13 m_1 \sigma^3 + \frac14 \lambda_\sigma \sigma^4.
\end{equation}
We suppose that pre-inflation multi-field dynamics dictate that inflation begins at the origin, $\sigma=0$, as 
in Raidal \etal\cite{DiChiara:2015euo}. This introduces only four parameters: four couplings in the potential --- 
the desired number of $e$-folds, \Nefold, is not an input parameter. We, furthermore, tune a 
dressed vacuum energy, 
$\rho$, such that the cosmological constant vanishes in the vacuum, \ie $V(\vev{\sigma},\ldots) + \rho = 0$.
Thus, low-scale inflation with $H \lll \MP$ is achieved provided $V(\sigma=0) = \rho \lll \MP^4$. This implies that the
potential must be fine-tuned such that $\left|V(\vev{\sigma})\right| \lll \MP^4$.  

The cosmological constant poses an infamous fine-tuning problem \see{Bousso:2007gp}. 
In almost all known models, agreement with measurements of the cosmological constant
requires extreme fine-tuning between a bare cosmological constant in the Lagrangian, $\rho_0$,
quantum corrections and contributions from spontaneous
symmetry breaking \ie $V(\vev{\sigma},\ldots)$. Because all models that we consider suffer from this fine-tuning problem, 
fine-tuning penalties from the cosmological constant would approximately cancel in ratios of
Bayesian evidences. We ensure that the second epoch of inflation cannot spoil the relaxion mechanism by 
applying conditions on the Hubble parameter during inflation.

\subsubsection{Relaxion physicality conditions}\label{Sec:Conditions}

There are parameter points for which the back-reaction to EWSB fails to trap the relaxion field in a 
minimum. If that were the case, the relaxion mechanism would fail and the point would be in 
severe disagreement with observations. Graham \etal list conditions
required for a successful relaxion mechanism\cite{Graham:2015cka}:
\begin{align}
H^2 \MP^2 &> \frac{\mu^2 m^2}{\kappa} & \describe{vacuum energy}\\
H &< m_b & \describe{barriers form}\\
H^3 &< m^2 \vev{a} & \describe{classical beats quantum}
\end{align}
We assign zero likelihood to a point that violates the resulting condition,
\begin{equation}\label{Eq:MassInequality}
\sqrt{\frac{\mu^2 m^2}{\kappa}} <  \MP \min(m_b, m^{2/3} \vev{a}^{1/3}).
\end{equation}
Graham \etal also list the conditions
\begin{align}
\Nefold &\gtrsim \frac{H^2}{\kappa \vev{a}^2} &\describe{inflation lasts long enough}\\
\kappa \vev{a} \mu^2 f &\sim m_b^3 \vev{h}  & \describe{barrier heights}
\end{align}
We assume that a first epoch of inflation is provided by the slow-rolling
relaxion fields, and cosmological constant later cancelled when the
Higgs and relaxion fields acquire VEVs, and that this epoch provides
an acceptable Hubble parameter, as described in \refcite{DiChiara:2015euo}. The latter condition is unnecessary as
we solve the potential with numerical methods, checking whether a solution exists.
The second epoch of inflation must, however, satisfy,
\begin{equation}
H < m_b
\end{equation}
to avoid destroying the periodic barriers.

\subsubsection{Calculation of (electroweak and QCD) observables}

We calculated the VEVs of the Higgs and relaxion fields with numerical methods based on
bisecting the interval in \refeq{Eq:int}, from which we calculated the mass of 
the $Z$-boson,
\begin{equation}\label{Eq:Predictions}
\mz = g  \vev{h}
\end{equation}
and \thetaQCD (see \refsec{Sec:ThetaQCD}). In the non-QCD relaxion model, \thetaQCD is an input 
parameter. The calculations for the inflationary observables were identical to those in the \SMsigma model.

\section{Bayesian analysis}\label{Sec:Results}

We calculated Bayesian evidences for our \SMsigma model and relaxion models with 
\code{(Py)-\allowbreak{}MultiNest}\cite{Buchner:2014nha,Feroz:2013hea,Feroz:2007kg,Feroz:2008xx}, which 
utilises the nested sampling algorithm\cite{Skilling:2004,skilling2006} 
for Monte-Carlo integration in Bayesian evidences in \refeq{Eq:Evidence} (though delta-functions were
first integrated by hand).\footnote{We utilised importance sampling, picked $1000$ live points and a stopping criteria of $0.01$ in \code{MultiNest}.} This requires two ingredients:
a likelihood function and a set of priors. Our likelihood function, summarised in \reftable{Table:Like}, was a product of at most five factors:
\begin{itemize}

\item \dataset{weak-scale}: A likelihood function for measurements of the mass of the $Z$-boson\cite{Agashe:2014kda}. 
In the SM, this is approximated by a delta-function and integrated by hand. In a relaxion model, this is impossible, 
as there is no analytic expression for the $Z$-boson mass as a function of the Lagrangian parameters.

\item \dataset{conditions}: If a relaxion model (\ie a point in a relaxion model's parameter space) 
violates physicality conditions in \refsec{Sec:Conditions}, we assign a likelihood of zero, since it would be in stark disagreement with observations.

\item \dataset{decay}: A likelihood function for the experimental lower-limit on $f_a$, the axion decay constant, approximated by a step-function \see{Dine:2007zp}.

\item \dataset{theta}: A likelihood function for the experimental upper-limit on \thetaQCD, approximated by a step-function\cite{Agashe:2014kda}. 

\item \dataset{inflation}: A likelihood for the spectral index, $n_s$, the ratio of scalar to tensor
perturbations, $r$, and the amplitude of scalar perturbations, $A_s$, 
from Planck and BICEP measurements\cite{Ade:2015tva,Ade:2015xua}. For simplicity,
we neglect correlations amongst Planck measurements and impose an
upper-limit for the scalar-to-tensor ratio.

\end{itemize}
We applied the likelihoods incrementally in five calculations per model: only \dataset{weak-scale};
adding \dataset{conditions}; adding a lower-bound on the axion decay constant, \dataset{decay}; 
adding an upper bound on \thetaQCD, \dataset{theta}; and finally adding BICEP/Planck measurements in \dataset{inflation}. This enabled us to assess the individual impacts
of the constraints.

\begin{table}[ht]
\centering
\begin{tabular}{ccc}
\toprule
Parameter & Measurement & Likelihood function\\
\midrule
\dataset{weak-scale}\\
\midrule
\mz & $91.1876 \pm 0.0021\gev$\cite{Agashe:2014kda} & Dirac in SM, Gaussian in relaxion\\
\midrule
\dataset{decay}\\
\midrule
 $f_a$ & $f_a \gtrsim 10^{9}\gev$\cite{Dine:2007zp} & Step-function\\
\midrule
\dataset{theta}\\
\midrule
\thetaQCD & $\thetaQCD \lesssim 10^{-10}$\cite{Agashe:2014kda} & Step-function\\
\midrule
\dataset{inflation}\\
\midrule
$r$   & $r < 0.12$ at $95\%$\cite{Ade:2015tva} & Step-function\\ 
$n_s$ & $0.9645 \pm 0.0049$\cite{Ade:2015xua} & Gaussian\\ 
$\ln(10^{10} A_s)$ & $3.094 \pm 0.034$\cite{Ade:2015xua} & Gaussian \\ 
\bottomrule
\end{tabular}
\caption{Likelihoods included in our Bayesian evidences for the scale of electroweak symmetry breaking, 
the axion decay constant, \thetaQCD and BICEP/Planck measurements of inflationary observables. Note that we neglect statistical correlations in 
Planck measurements of inflationary observables.}
\label{Table:Like}
\end{table}

We picked uninformative scale-invariant priors for the dimensionful Lagrangian parameters and cut-off because we
are ignorant of their scale, a linear prior for \thetaQCD, reflecting a shift-symmetry, and a linear 
prior for \Nefold because the number of $e$-folds is already a logarithmic quantity. 
Our prior ranges are summarised in \reftable{Table:Priors}. All massive parameters --- $\mu^2$ and $m^2$ and inflaton masses --- 
receive quadratic corrections from a cut-off, such that we expect that without fine-tuning 
$\mu^2 \sim m^2 \sim \Lambda^2$. The main difference between the priors for our QCD relaxion model and 
general relaxion model is that in the former,
the barrier height is related to the QCD scale, whilst in the latter, it is no 
greater than about the weak scale. 

\mxnewcommand{\dash}{\text{, }}
\afterpage{\clearpage}  
\begin{table}[p]
\centering
\footnotesize
\begin{tabular}{ccc}
\toprule
Parameter & Prior & \\
\midrule
\SMsigma\\
\midrule
$\mu^2$ & Log & $10^{-40} \dash 1$\\
$\lambda$ & Log & $10^{-4} \dash 4\pi$\\
$\Lambda^2$ & Log & $10^{-4} \dash 1$\\
\midrule
$m_\sigma^2$ & Log & $10^{-40} \dash 1$\\
$\lambda_\sigma$ & Log & $10^{-20} \dash 4\pi$\\
\Nefold & Linear & $50 \dash 500$ \\ 
\midrule
\thetaQCD & Linear & $0, \pi$ \\
\midrule
QCD relaxion\\
\midrule
$\mu^2$ & Log & $10^{-40} \dash 1$\\
$\lambda$ & Log & $10^{-4} \dash 4\pi$\\
$\Lambda^2$ & Log & $10^{-4} \dash 1$\\
$\kappa$ & Log & $10^{-4} \dash 4\pi$\\
$\vev{a}$ & Log & $10^{-20} \dash 1$\\
$m_b$ & Log & $10^{-1} \Lambda_\text{QCD} \dash 10\Lambda_\text{QCD}$\\
$m^2$ & Log & $10^{-60} \dash 1$\\
$f$ & Log & $10^{-20} \dash 1$\\
\midrule
$m_1$ & Log & $10^{-100} \dash 1$\\
$m_2^2$ & Log & $10^{-100} \dash 1$\\
$m_3^3$ & Log & $10^{-100} \dash 1$\\
$\lambda_\sigma$ & Log & $10^{-40} \dash 4\pi$\\
\midrule
Non-QCD relaxion, as for QCD relaxion except\\
\midrule
$m_b$ & Log & $10^{-6} \vev{h} \dash 10^{-1} \vev{h}$\\
\thetaQCD & Linear & $0, \pi$ \\
\bottomrule
\end{tabular}
\caption{Priors for parameters in SM augmented with scalar-field inflation (\SMsigma) and 
relaxion models. Masses are in Planck units.}
\label{Table:Priors}
\end{table}

\subsection{Evidences}

The evidences and Bayes-factors for the \SMsigma and relaxion models are summarised 
in \reftable{Table:Evidences}. We find that, considering only
a measurement of the weak scale (\ie \dataset{weak-scale}), relaxion models are favoured by colossal
Bayes-factors of about $10^{30}$. This is similar to findings for the constrained minimal supersymmetric SM
versus the SM\cite{Fowlie:2014xha}, and was expected, as the SM with a Planck-scale
cut-off makes an egregious generic prediction for the weak scale.

The physicality conditions (\dataset{conditions} in \refsec{Sec:Conditions}) 
dramatically impact the preference for relaxion models. The conditions wipe-out a fraction of the 
relaxion models' parameter spaces and shrink the
Bayes-factors by about $10^{-28}$. The preference for relaxion models versus the
SM almost entirely disappears. In other words, despite
their success in solving the hierarchy problem, relaxion models are hamstrung
by severe fine-tuning associated with their physicality conditions. 

The preference for the QCD relaxion model is further damaged by measurements of
the axion decay constant, \dataset{decay}, and the \thetaQCD, \dataset{theta}.
The latter results in approximately zero preference for the QCD relaxion model as it 
predicts that $\thetaQCD \approx \pi/2$ (see \refeq{Eq:ThetaQCD}). The preference of about $10^2$ for a non-QCD relaxion model versus the SM is unaffected by
\dataset{decay} and \dataset{theta}. 

The final data-set of inflationary observables (\dataset{inflation}) 
is the nail in the coffin for the relaxion models that we consider. Low-scale inflation, required in the relaxion paradigm, 
suffers severe fine-tuning as it requires a light scalar, and thus results in partial Bayes-factors of about $10^{-30}$ for 
relaxion models versus the SM\@. Thus, all data considered, the \SMsigma model is favoured by
a Bayes-factor of at least about $10^{25}$.

We note that \refeq{Eq:MassInequality} results in an approximate limit of $\mu^2 \sim m^2 \lesssim (10^8\gev)^2$, such that
by chance
\begin{equation}\label{Eq:LimitApprox}
\frac{\mu^2}{\beta \MP^2} \frac{m^2}{\beta \MP^2} \sim \frac{\mz^2}{\beta \MP^2}.
\end{equation}
The factors are in fact approximately the
fractions of parameter space in which a scalar mass is fine-tuned to be so light versus a 
cut-off, $\MP$. Thus tuning two scalar masses --- $\mu^2$ and $m^2$ --- in a relaxion 
model to be $\mu^2 \sim m^2 \sim (10^8 \gev)^2$ results in a similar fine-tuning
penalty as tuning a single scalar mass such that $\mz \sim 100\gev$. This, in
essence, explains why the evidence for the SM and relaxion models are similar, if one 
considers only \dataset{weak + conditions}. Note that lowering the quadratic corrections
by supersymmetrizing the SM and relaxion models \see{Batell:2015fma} could favour
relaxion models, as from \refeq{Eq:LimitApprox} a Bayes-factor might scale as the cut-off squared.
Lowering the Planck mass, on the other hand,
might help slightly less, as it would lower the bounds on scalar masses from \refeq{Eq:MassInequality}.

\begin{figure}
\centering
\subfloat[][SM]{\label{Fig:SM_Z}
\includegraphics[width=0.49\textwidth]{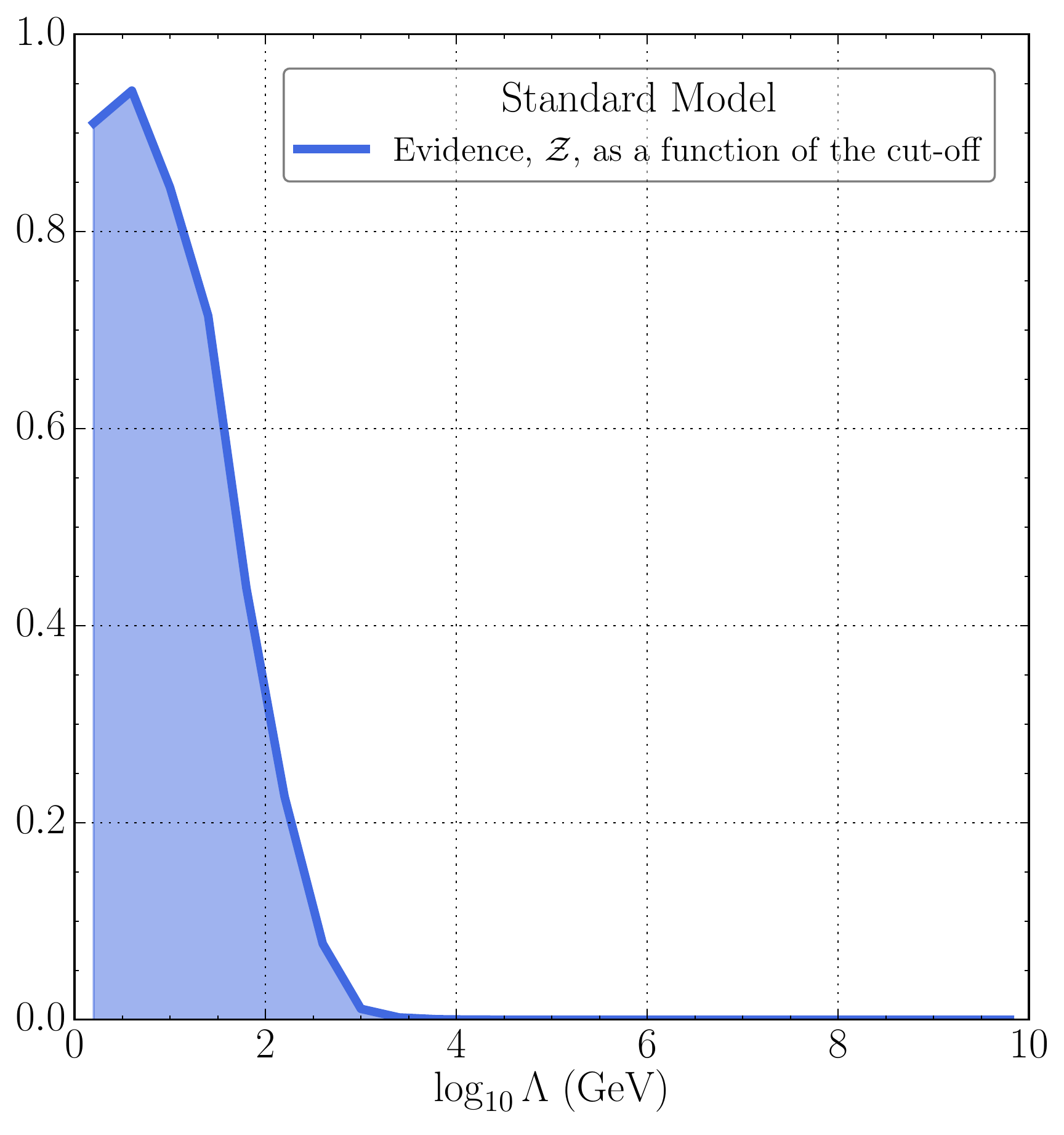}
}
\centering
\subfloat[][Relaxion]{\label{Fig:relaxion_Z}
\includegraphics[width=0.49\textwidth]{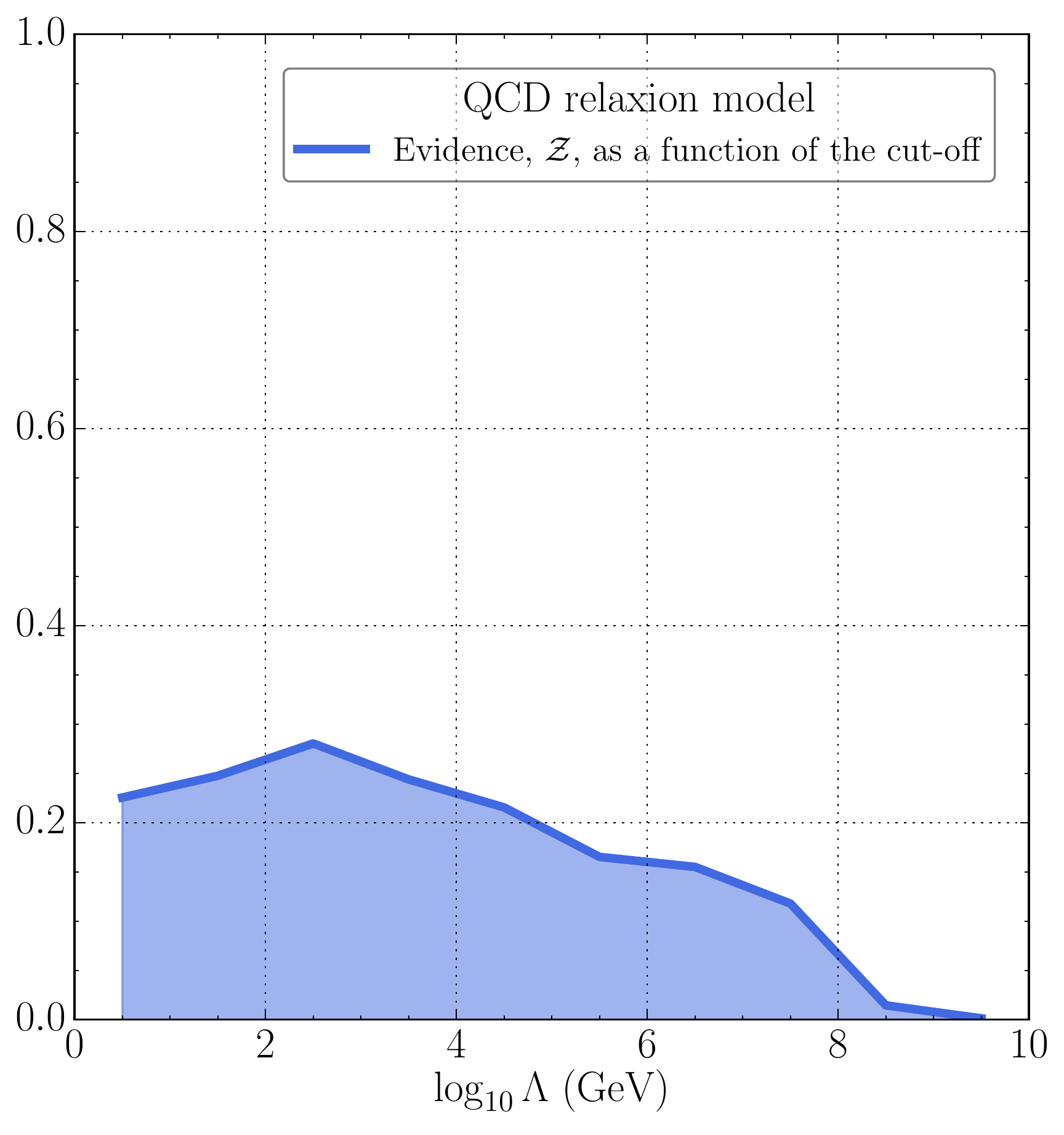}
}
\caption{The evidence as a function of the cut-off, $\Lambda$, in \protect\subref{Fig:SM_Z} the SM
and \protect\subref{Fig:relaxion_Z} a relaxion model. The evidence includes \dataset{weak-scale} and 
\dataset{conditions}. This illustrates that a relaxion model could be significantly favoured if the 
cut-off were lowered from the Planck scale to about $10^8 \gev$ \eg by supersymmetrizing
the SM and relaxion model. The evidences are plotted in arbitrary units.}   
\label{Fig:Z}
\end{figure}

To further investigate this issue, we relaxed the Planck-scale cut-off, plotting evidence
as a function of the cut-off in the SM and our QCD relaxion model in \reffig{Fig:Z}. By doing so, we wish to confirm
that our QCD relaxion model would be favoured versus the SM, were the cut-off much lower than the Planck scale. 
We find in \reffig{Fig:Z} that, although we previously found that the relaxion model was not favoured versus the SM with a Planck-scale
cut-off, if the cut-off were lowered in each model to about $10^2 \gev \lesssim \Lambda\lesssim 10^8 \gev$, 
the relaxion model could be significantly favoured. In other words, 
the relaxion mechanism may solve the little-hierarchy problem in a supersymmetric model, 
but not the hierarchy problem by itself. By itself, our QCD relaxion model cannot improve fine-tuning 
compared to the SM\@. 

With a cut-off allowed to be as low as $10\tev$, 
considering \dataset{weak-scale}, \dataset{conditions} and \dataset{decay}, the Bayes-factor favours our QCD relaxion model 
by $10^6$ versus the SM and about $10^{30}$ versus the SM with Planck-scale quadratic corrections. 
Including low-scale inflation in \eg a supersymmetrized relaxion model, however, 
might necessitate an inflaton mass $m_\sigma \ll M_\text{SUSY}$. This little-hierarchy
problem could scotch the Bayes-factor of $10^6$ in favour of the supersymmetrized relaxion model.

\definecolor{RoyalBlue}{HTML}{4169E1}
\newcommand{\cc}{\cellcolor{RoyalBlue!50}}
\newcommand{\invgev}{\ensuremath{\,\cdot\,\text{GeV}}}
\begin{table}[ht]
\centering
\footnotesize
\begin{tabular}{cccccc}
\toprule
Data-set & \dataset{weak-scale} & \dataset{+=conditions} & \dataset{+=decay} & \dataset{+=theta} & \dataset{+=inflation}\\
\midrule
\ev(\SMsigma)\invgev & $10^{-34}$ & & & $10^{-45}$ & $10^{-53}$\\
\ev(\text{relaxion})\invgev & $10^{-4}$ & $10^{-32}$ & & $10^{-43}$ & $10^{-78}$\\
\ev(\text{QCD relaxion})\invgev & $10^{-4}$ & $10^{-34}$ & $10^{-39}$ & $\lll 10^{-39}$ & $\lll 10^{-80}$\\
\midrule
$B(\text{relaxion}/\SMsigma)$ & \cc $10^{30}$ & $10^{2}$ & & $10^{2}$ & \cc $10^{-25}$\\
$B(\text{QCD relaxion}/\SMsigma)$ & \cc $10^{30}$ & $1$ & $10^{-5}$ & $\lll 10^{6}$ & \cc $\lll 10^{-27}$\\
$B(\text{QCD relaxion}/\text{relaxion})$ & $1$ & $10^{-2}$ & $10^{-7}$ & $\lll 10^{4}$ & $\lll 10^{-2}$\\
\midrule
$P(\text{relaxion}/\SMsigma)$ & & $10^{-28}$ & & $1$ & $10^{-27}$\\
$P(\text{QCD relaxion}/\SMsigma)$ & & $10^{-29}$ & $10^{-5}$ & $\lll 1$ & $10^{-33}$\\
$P(\text{QCD relaxion}/\text{relaxion})$ & & $10^{-2}$ & $10^{-5}$ & $\lll 1$ & $10^{-6}$\\
\bottomrule
\end{tabular}
\caption[foo bar]{
Evidences, \ev, Bayes-factors, $B$ and partial Bayes-factors, $P$, for the SM augmented with scalar-field inflation (\SMsigma), a relaxion toy-model and a QCD relaxion toy-model. We apply data incrementally in five data-sets: 
\begin{enumerate*}[label=\itshape(\roman*)]
\item the $Z$-boson mass (\dataset{weak-scale}), 
\item physicality conditions in relaxion models (\dataset{conditions}), 
\item constraints on the axion decay constant (\dataset{decay}),
\item \thetaQCD (\dataset{theta}) and 
\item BICEP/Planck measurements of inflationary observables (\dataset{inflation}). 
\end{enumerate*}
A Bayes-factor is a ratio of evidences, indicating the change in relative plausibility of two models in light of all data considered thus far. A partial Bayes-factor is a ratio of Bayes-factors, indicating the change in relative plausibility of two models in light of incrementing the data by a single data-set. A ratio of greater than one indicates that a relaxion toy-model is favoured.
We highlight our most important findings in blue: that relaxion toy-models are favoured by about $10^{30}$ by the $Z$-boson mass, but that once all constaints are included, that preference is reversed to about $10^{-25}$ against relaxion toy-models.	
}
\label{Table:Evidences}
\end{table}

\subsection{Observables}

\begin{figure}
\centering
\subfloat[][SM]{\label{Fig:SM_MZ}
\includegraphics[width=0.49\textwidth]{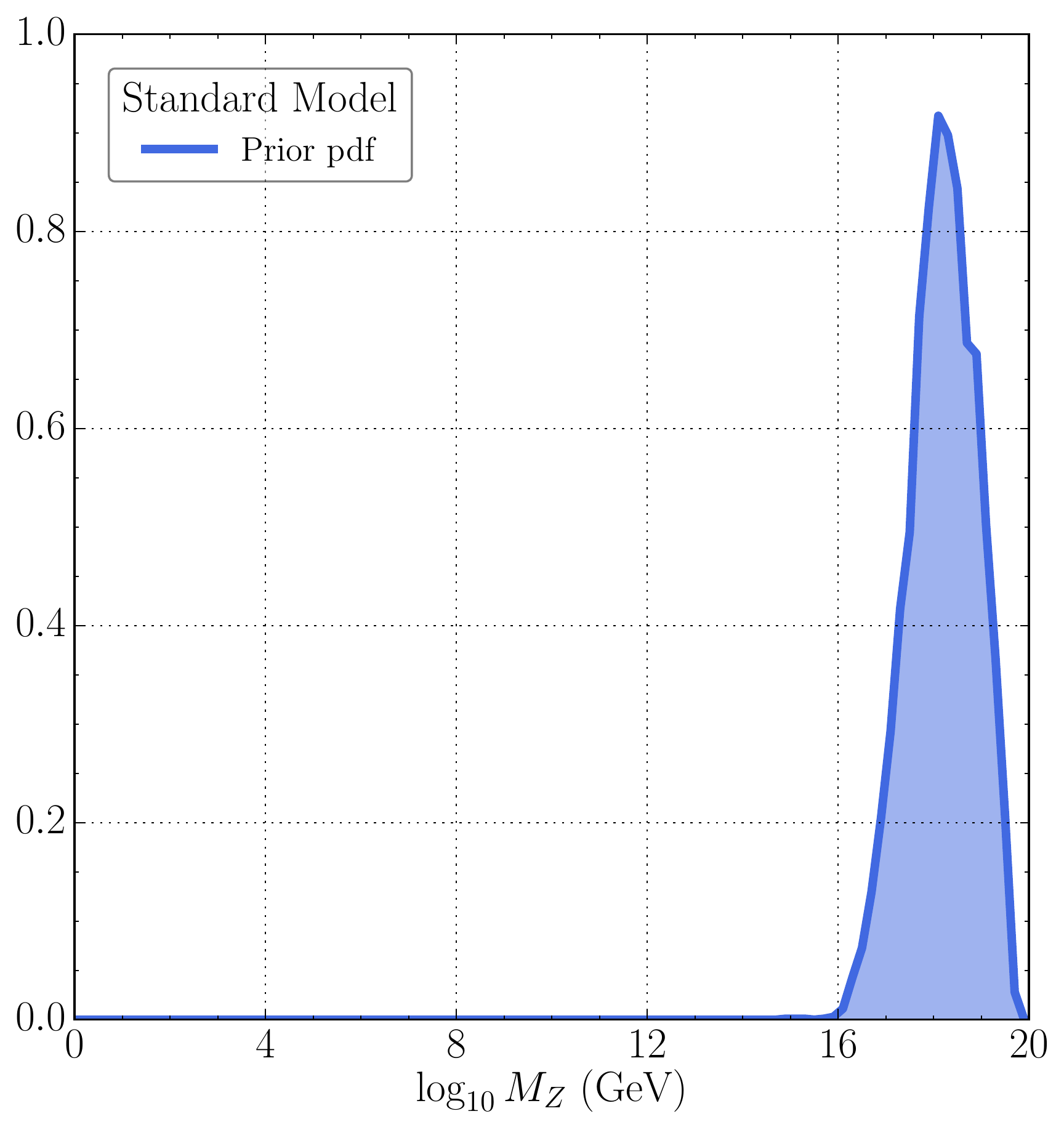}
}
\centering
\subfloat[][Relaxion]{\label{Fig:relaxion_MZ}
\includegraphics[width=0.49\textwidth]{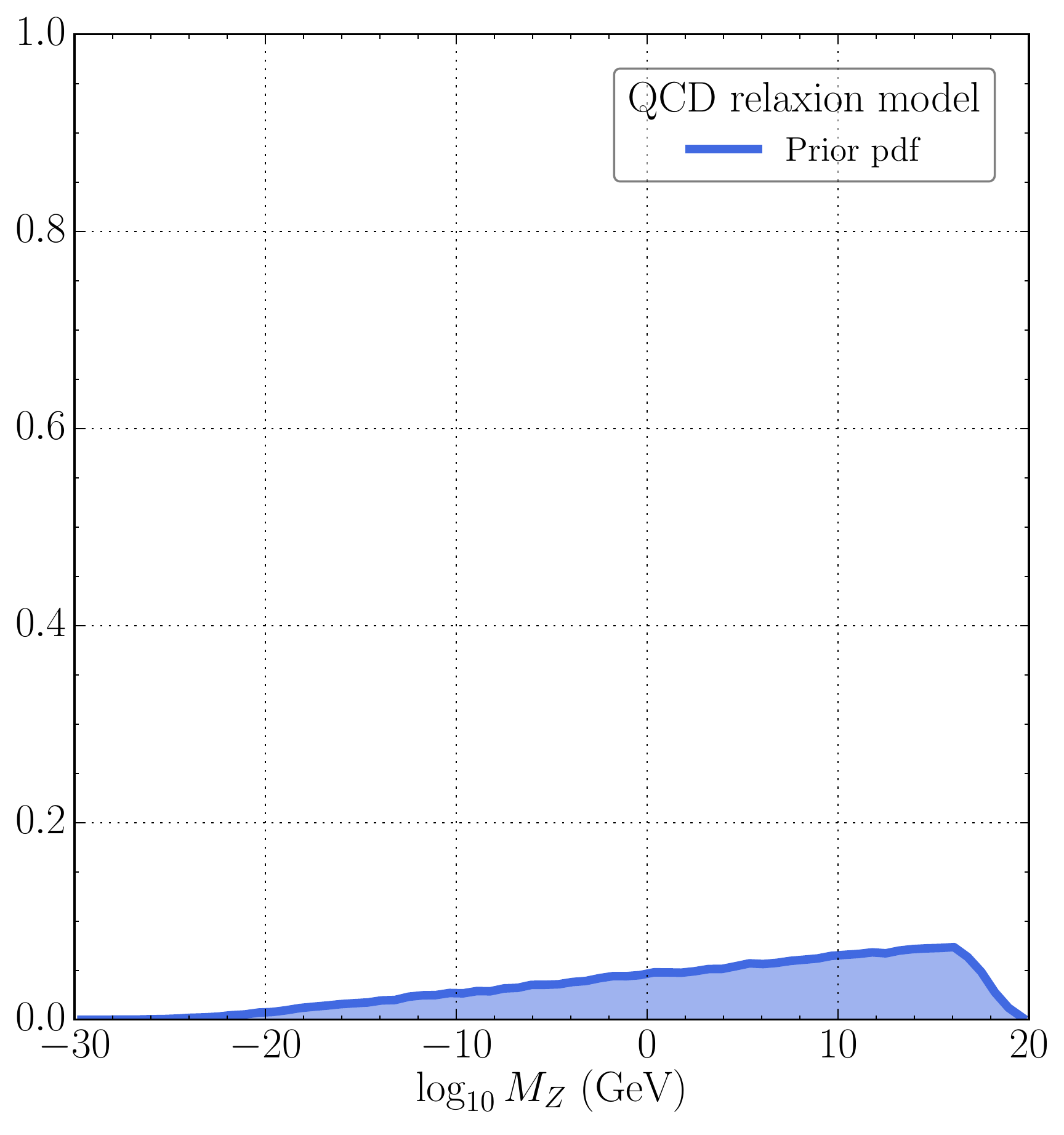}
}
\caption{Prior distribution of $\log_{10}$ of the $Z$-boson mass in \protect\subref{Fig:SM_MZ} the SM
and \protect\subref{Fig:relaxion_MZ} a QCD relaxion model including no data. 
The density at the correct weak scale in the relaxion model is much
greater than that in the SM\@. This
illustrates that the relaxion mechanism improves fine-tuning of the weak scale
with respect to the SM\@. The densities are plotted in arbitrary units.}   
\label{Fig:MZ}
\end{figure}

To illustrate the resolution of the hierarchy problem, in \reffig{Fig:MZ}
we plot the priors for the $Z$-boson mass in the SM and our QCD relaxion model that result from
the non-informative priors for Lagrangian parameters in \reftable{Table:Priors}, that is,
\begin{equation}
\pg{\log\mz}{\model} = \int \delta(\log\mz - \log\mz(\params))\,\pg{\params}{\model} \prod \dif \params.
\end{equation}
This illustrates their generic predictions for the weak scale. This would be 
numerically equivalent to the Bayesian evidence if our data were $\log\mz$ and 
we approximated our measurement with a Dirac function. Whereas the SM favours a 
weak scale close to the Planck scale, the relaxion model results in 
considerable probability mass at scales much less than the Planck scale, 
resolving 
the hierarchy problem. We find that if the relaxion is the QCD axion,
the posterior probability that $\thetaQCD \lesssim 10^{-10}$ is negligible,
confirming our expectations.

\section{Discussion and conclusions}\label{Sec:Conclusions}

We constructed models that utilised a relaxation mechanism recently
proposed by Graham \etal to solve the hierarchy problem. Unfortunately, finding the weak scale
in relaxion models involves solving a transcendental equation with numerical methods. 
We presented an analytic expression for an interval bounding the weak scale and
an analytic expression for a lower bound on \thetaQCD, confirming that $\thetaQCD\approx\pi/2$ 
if the relaxion is the QCD axion.

We performed the first statistical analysis of a relaxion
model by scanning relaxion models' parameter spaces with the nested sampling algorithm, considering
data from measurements of the weak scale, the axion decay constant, \thetaQCD
 and BICEP/Planck measurements of inflationary
observables $r$, $n_s$ and $A_s$. This resulted in so-called Bayesian evidences 
for our relaxion models augmented with scalar-field inflation. In a similar manner, we calculated Bayesian evidences
for the SM augmented with scalar-field inflation.

We found that the Bayes-factors --- ratios of Bayesian evidences that indicate
how one ought to update one's relative prior belief in two models --- favoured
relaxion models versus the SM by a colossal factor of about $10^{30}$ 
if one considered only the weak scale. Once we included physicality conditions upon 
inflation during relaxation, however, the Bayes-factors were decimated to about
$100$ for the non-QCD relaxion model and about $1$ for the QCD relaxion model.

Constraints upon the QCD relaxion decay constant and \thetaQCD shatter faith in the 
QCD relaxion model (in the parlance of conventional frequentist statistics, the model is excluded). 
Finally, inflationary observables measured by BICEP/Planck demolish the plausibility
of the surviving, non-QCD relaxion model as the SM augmented with scalar-field inflation
is favoured by a Bayes-factor of about $10^{25}$. This stems from a constraint upon the Hubble 
parameter during inflation; $H\lll \MP$ must be fine-tuned such that inflation cannot destroy periodic 
barriers in the relaxion potential. 

Thus, whilst the analysed relaxion models indeed solve the hierarchy problem leading to Bayes-factors of about 
$10^{30}$ in their favour, the same Bayes-factors are scotched by 
constraints upon parameters in the relaxion potential and the Hubble parameter during
inflation, ultimately leading to a Bayes-factor of about $10^{25}$ in favour of the SM augmented with 
scalar-field inflation. We anticipate, furthermore, that detailed consideration of baryogenesis 
and thermal effects (including the disastrous possibility of reheating restoring electroweak symmetry)
would further damage the plausibility
of relaxion models and conclude that the required unconventional cosmology is the Achilles' heel of the relaxion mechanism.
If the associated problems were overcome, a relaxion model would be vastly more plausible 
than the SM; however, we know of no such model.

\section*{Acknowledgements}
We thank Kristjan Kannike for helpful discussions. LM and MR are supported by the grants IUT23-6, PUTJD110 and by EU through the ERDF CoE program. 
AF, GW and CB are in part supported by the ARC Centre of Excellence for Particle Physics at the Tera-scale.

\end{document}